\documentclass[prc,twocolumn,showpacs,showkeys,superscriptaddress,nofootinbib]{revtex4}
\usepackage{graphicx}
\def\be{\begin{eqnarray}}
\def\ee{\end{eqnarray}}
\def\bc{\begin{center}}
\def\ec{\end{center}}

\newcommand{\lsim}{\stackrel{\scriptstyle <}{\phantom{}_{\sim}}}
\newcommand{\gsim}{\stackrel{\scriptstyle >}{\phantom{}_{\sim}}}
\begin{document}
\title{Hydrodynamical description of  a hadron-quark first-order  phase
transition.}
\author{V.V. Skokov}
\affiliation{GSI, Plankstra\ss{}e 1, D-64291 Darmstadt, Germany}
%\thanks{present address}
\affiliation{JINR, 141980 Dubna, Moscow Region, Russia}
\author{D.N. Voskresensky}
\affiliation{GSI, Plankstra\ss{}e 1, D-64291 Darmstadt, Germany}
\affiliation{MEPhI, Kashirskoe
  Avenue 31, RU-11549 Moscow, Russia}
%=================================================================
\begin{abstract}
Solutions of hydrodynamical equations are presented for the
equation of state of the Var der Waals type allowing for the first
order phase transition. Attention is focused on  description of
the hadron-quark phase transition in heavy ion collisions. It is
shown that fluctuations dissolve and grow as if the fluid is
effectively very viscous. Even in spinodal region germs are
growing  slowly due to viscosity and critical slowing down. This
prevents enhancement of fluctuations in the near-critical region,
which is frequently considered as a signal of the critical point
in heavy ion collisions.
\end{abstract}
\date{\today}

\maketitle

There are many phenomena, where first-order phase transitions
occur between phases with different densities.  Description of
such phenomena should be similar to that for the gas-liquid phase
transition.
Thereby it is worthwhile to find corresponding solutions of
hydrodynamical equations. Though some simplified analytical
\cite{PS,MSTV90} and fragmentary two-dimensional numerical
solutions \cite{Onuki} have been found, many problems remain
unsolved. In nuclear physics different first-order phase
transitions (e.g., to  pion, kaon condensates and to the quark
state)  may occur in neutron stars \cite{MSTV90,Glendenning} and
in heavy ion collisions \cite{Randrup,Shuryak:2008eq}. At low
energies gas-liquid transition occurs \cite{Randrup}.  It is also
expected that at finite baryon density the hadron -- quark gluon
plasma (QGP) phase transition, which  might  manifest itself in
violent nucleus-nucleus collisions, is of the first-order
\cite{Shuryak:2008eq}.
 The hydrodynamical approach is efficient for
description  of heavy-ion collisions in a broad  energy range
(e.g. see \cite{SG,Romatschke:2007mq,Shuryak:2008eq}).

In this letter the dynamics of a first-order phase transition is
described  by equations of non-ideal non-relativistic
hydrodynamics: the  Navier-Stokes equation, the continuity
equation, and  general equation for the heat transport. We solve
these equations numerically in two spatial dimensions, $d=2$, and
analytically for arbitrary $d$  in the vicinity of the critical
point. Then we perform estimations for the case of the hadron --
QGP  transition.

 The
best known example to illustrate principal features of a first
order phase transition is the Van der Waals fluid. The pressure is
given by
$P_{\rm VW}[n,T]={nT}/{(1-bn)}-n^2 a ,$
  where $T$ is the temperature, $n$ is the density of a conserving charge (e.g., the baryon charge),
   parameter $a$ governs the strength of a mean field attraction
and $b$
 controls a short-range repulsion.
In practice we use a modified
 Van der Waals (mVW) equation of state (EoS):
$P[n,T]=f(T)P_{\rm VW}[n,T],\,$ where the function $f(T)$ is
chosen so that $\frac{\partial P}{\partial
T}|_{(n_{cr},T_{cr})}=0$ at the critical density $n_{cr}$ and the
critical temperature $T_{cr}$. We use $f(T)\simeq C(\delta{\cal{
T}})\left[1- 2 (\delta \cal{T})\right]^2 ,$
 $\delta
{\cal{T}}=(T-T_{cr})/T_{cr}$, and the  pre-factor $C(\delta
{\cal{T}})=[1+4(\delta {\cal{T}})^2]^{-1}$ is chosen to reproduce
ideal gas EoS
  for sufficiently low $n$ and high $T$.
This modification allows us to parameterize the EoS with two
minima in the free energy, being convenient for analytical
treatment of the problem near the critical point
$(n_{cr},T_{cr})$. We expand the quantities entering EoS and
equations of hydrodynamics near a reference point ($\rho_{\rm r}
,T_{\rm r}$) chosen somewhere in the vicinity of the critical
point on the plane $P(\rho ,T)$, where $\rho =mn$ is the mass
density, $m$ is the  mass of the constituent.
Assuming smallness of the velocity $\vec{u}(\vec{r},\tau )$ of the
germ we linearize hydrodynamical equations in  $u$, density
$\delta\rho =\rho -\rho_{\rm r}$ and  temperature $\delta
T=T-T_{\rm r}$.
 Applying then
operator "$\mbox{div}$" to  the Navier-Stokes equation and taking
$z=\mbox{div}\,\vec{u}$ from the continuity equation we obtain
\cite{PS,MSTV90}:
 \be\label{v-t}
\frac{\partial^2 \delta\rho}{\partial t^2}=\Delta \left[\delta P
+\rho_{\rm r}^{-1}\left(\tilde{d}\eta_{\rm r} + \zeta_{\rm r}
\right)\frac{\partial \delta\rho}{\partial t} \right],
 \ee
$\tilde{d}= {2(d-1)}/{d}$. The pressure $ \delta P=P-P[\rho_{\rm
r},T_{\rm r}]=
 \rho_{\rm r}\frac{\delta [F(\delta \rho ,T)]}{\delta (\delta \rho)}|_{T}$ is expressed through the
Helmholtz free energy $F$ for slightly inhomogeneous
configurations;
 $\eta_{\rm r}$ and $\zeta_{\rm r}$ are the first
(shear) and second (bulk) viscosities; $ \Delta =\partial_{x_1}^2
+....+\partial_{x_d}^2 .$

Note that {\em{thus derived Eq. (\ref{v-t})  differs from the
phenomenological Landau equation}} for the nonconserving order
parameter $\partial_{t}\phi =-\gamma ({\delta F}/{\delta \phi}),$
$\gamma =const$,  and from equations used for the description of
the dynamics of  first-order phase transitions in heavy ion
collisions \cite{Gavin} and in relativistic astrophysical problems
\cite{astro}. The difference with the  Landau  equation
disappears, if one sets zero the square bracketed term in the
r.h.s. of Eq. (\ref{v-t}). From the first glance, such a procedure
is legitimate, if space-time gradients are small. However for a
germ, being prepared in a fluctuation at $t=0$ with a distribution
$\delta\rho (t=0,\vec{r})=\delta\rho (0,\vec{r})$,
 the condition ${\partial \delta\rho
(t, \vec{r}) }/{\partial t}|_{t=0}\simeq 0$ should also be
fulfilled (otherwise there appears a  kinetic energy term). Two
initial conditions cannot be simultaneously fulfilled, if the
equation contains  time derivatives of the first-order only.
 Thus,  {\em{there exists an initial
stage of the dynamics of  phase transitions ($t\lsim t_{\rm
init})$, which is not described by the standard Landau equation.}}

For low velocities the heat transport is described by the heat
conductivity equation $ c_{V}\frac{\partial  T}{\partial t}
=\kappa \Delta  T ,$ where $\kappa$ is the heat conductivity and
$c_V$ is the specific heat. Time scale  of the temperature
relaxation is $ t_T = R^2 (t_T ) c_{V} /\kappa ,$ where $R(t)$ is
the size of the germ. On the other hand, time scale of the density
relaxation, following Eq. (\ref{v-t}), is $t_{\rho}\propto R$ (we
show below that a germ of rather large size grows with constant
velocity). Evolution of the germ is governed by the slowest mode.
When  sizes of germs begin to exceed the value $R_{\rm fog}$,
where $R_{\rm fog}$ is  the size at which  $t_T =t_{\rho}$, the
growth is slown down. Thus number of germs with the size $R\sim
R_{\rm fog}$ grows with  time and {\em{there appears a metastable
state called the fog.}}

 For further convenience we  choose $\rho_{\rm r} =\rho_{cr}$, $T_{\rm r}
=T_{cr}$ and expand the Helmholtz free energy in $\delta\rho$ and
$\delta{\cal{ T}}$:
 \be\label{fren}
\delta F = \int \frac{d^3 x}{\rho_{\rm r} }\left[ \frac{c[\nabla
(\delta \rho)] ^2}{2}+\frac{\lambda (\delta
\rho)^4}{4}-\frac{\lambda v^2 (\delta \rho) ^2}{2}-\epsilon \delta
\rho \right],
 \ee
$\delta F = F[\rho ,T] -F [\rho_{\rm r},T_{\rm r}]$.
Then
$ a =\frac{9}{8}\frac{T_{cr}}{n_{cr}},\,$ $
b=\frac{1}{3n_{cr}},\,$ $ v^2 =-\frac{3m^2 T_{cr}\delta{\cal{
T}}}{2ab}=4|\delta{\cal{T}}|n_{cr}^2 m^2 ,$
$ \lambda =\frac{3ab}{2m^3}=\frac{9}{16}\frac{T_{cr}}{n_{cr}^2
 m^3},\,$
$  \epsilon = \frac{27(T_{cr}\delta {\cal{
T}})^2}{4a}=6n_{cr}T_{cr}(\delta{\cal{T}})^2.$
Introducing dimensionless variables $\delta \rho =v \psi$, $\xi_i
=x_i /l$, $i =1 ,\cdots , d$, ${\tau}=t/t_0$,  we arrive at
equation
 \be\label{dimens}
 &&- \beta \frac{\partial^2 \psi }{\partial
{\tau}^2} =\Delta_{\xi}\left(\Delta_{\xi}\psi +2\psi
 (1-\psi^2)+\widetilde{\epsilon}- \frac{\partial \psi}{\partial
 {\tau}}\right),\\
 &&l=\left({2c}/({\lambda v^2})\right)^{1/2} ,\,\, t_0
 ={2(\tilde{d}\eta_{\rm r} +\zeta_{\rm r} )}/({\lambda v^2
 \rho_{\rm r}}),\,\,\nonumber\\
 &&\widetilde{\epsilon}={2\epsilon}/({\lambda v^3})=(8/3) |\delta{\cal{T}}|^{1/2},\,
  \beta
 =c\rho_{\rm r}^2 / [ \tilde{d}\eta_{\rm r}
 +\zeta_{\rm r}
 ]^{2}.\nonumber
  \ee
Thus $l\propto |\delta{\cal{ T}}|^{-1/2}$ and $t_0\propto
 |\delta{\cal{ T}}|^{-1}$.

There exists an opinion, cf.   Ref.  \cite{Stephanov},
 that, if at some incident energy the trajectory passes in the vicinity of the critical point,
 the system may  linger longer in this region due to divergence of susceptibilities
 that may reflect on observables.
 Contrary, we argue that  {\em{ fluctuational effects
 in the vicinity of the critical point in heavy ion collisions can  hardly  be pronounced}},
 since all relevant  processes are proved to be frozen for $\delta T\rightarrow 0$, while the system
 passes this region during a finite time.

 To describe configurations of different symmetry we search two-phase solution of Eq.
(\ref{dimens}) in the form
 \cite{PS,MSTV90},
 \be\label{sol}
 \psi =\mp \tanh[\xi -\xi_0
 ({\tau})]+\widetilde{\epsilon}/4 ,
  \ee
 $\xi =\sqrt{\xi^2_1
+\xi_2^2 +\xi_3^2}$ for droplets/bubbles  ($d_{\rm sol}=3$), $\xi
=\sqrt{\xi^2_1 +\xi_2^2}$ for rods ($d_{\rm sol}=2$)  and  $\xi
=\xi_1 =x/l$ for kinks ($d_{\rm sol}=1$)  in $d=3$ space.
 For $\widetilde{\epsilon} >0$ upper sign
solution describes evolution of droplets (or rods and kinks of
liquid phase) in a metastable super-cooled vapor medium.
 The
lower sign solution circumscribes  then   bubbles (or kinks and
rods of gas phase) in a stable liquid medium.

The boundary layer has the length $|\xi - \xi_0 (\tau)|\sim 1$.
Outside this layer corrections to homogeneous solutions are
exponentially small.
 Considering motion of the boundary for $\xi_0
(\tau)\gg 1$ we may put $\xi \simeq
 \xi_0 (\tau)$ in (\ref{sol}).
Then keeping only linear terms in $\widetilde{\epsilon}$ in  Eq.
(\ref{dimens}), we arrive at equation  for $\xi_0 (\tau)$:
 \be\label{ks}
\frac{\beta}{2}\frac{d^2\xi_0}{d\tau^2} =
\pm\frac{3}{2}\widetilde{\epsilon}-\frac{d_{\rm sol}-1}{ \xi_0
(\tau)}-\frac{d\xi_0}{d\tau}.
 \ee

Substituting  (\ref{sol}) in (\ref{fren})  we obtain
 \be\label{Fsol}
&&\delta F [\xi_{0}] =\frac{2\pi^{3/2}\Lambda^{3-d_{\rm
sol}}\lambda v^4 l^{d_{\rm sol}}}{\Gamma(d_{\rm sol}/2)\Gamma
(1+(3-d_{\rm sol})/2)\rho_{\rm r}} \\ &&\times\left[ \mp
\widetilde{\epsilon} \xi_{0}^{d_{\rm sol}}/{d_{\rm sol}}
+2\xi_{0}^{d_{\rm sol}-1}/3 \right], \nonumber
  \ee
 $2\Lambda$ is the diameter, height of cylinder and the length of the squared plate
 for $d_{\rm sol}=3,2$ and $1$, respectively;
 $\Gamma$ is the Euler $\Gamma$-function.
The first term in (\ref{Fsol}) is the volume term and the second
one is the surface contribution, $\delta  F_{\rm surf}$. At fixed
volume in $d =3$ space, the surface contribution  for
droplets/bubbles is
 smaller than for rods and slabs. Thereby if  a germ prepared in a
 fluctuation  is initially nonspherical  it   acquires spherical form
 with
passage of time.
Surface term is $\delta  F_{\rm surf} \equiv \sigma S$, $S$ is the
surface of the germ, $\sigma$ is the surface tension, and  the
gradient term in (\ref{fren}) is then $\delta F_{\rm surf}^{\rm
grad}
 = \frac{2 v^2
c}{3l\rho_{\rm r}}S=\frac{1}{2}\delta  F_{\rm surf}.$ Thus we are
able to find relations: $\sigma = \sigma_0 |\delta
{\cal{T}}|^{3/2},$ ${\sigma_0^2}= {32mn_{cr}^2 T_{cr}}c;$
$l=\frac{\sigma_0}{6T_{cr}n_{cr}|\delta{\cal{T}}|^{1/2}}.$ There
are two dimensionless parameters in  (\ref{dimens}) and
(\ref{ks}): $\widetilde{\epsilon}$ and $\beta$. The value
$\widetilde{\epsilon}$ distinguishes metastable and stable state
minima in the free energy,
 $
\beta
 = (32T_{cr})^{-1}[\tilde{d}\eta_{\rm r} +\zeta_{\rm r} ]^{-2}\sigma_0^2 m
  $
 controls dynamics.
{\em{The larger viscosity and the smaller surface tension, the
effectively more viscous is the fluidity of germs.}} For $\beta
\ll 1$ one deals with effectively viscous fluid and at $\beta \gg
1$, with perfect fluid.

At hand of  Eq. (\ref{ks}) consider analytically several typical
solutions for germ evolution. Consider evolution of germs of
stable phase in metastable matter.

1) {\em{Short time evolution of a germ.}} For small $\tau$
(initial stage) using Taylor expansion in $\tau$ and assuming zero
initial velocity, $\frac{d\xi_0 }{d\tau}|_{\tau =0}\simeq 0$, we
obtain
 $$R(t)\simeq R_0 +({w t^2}/{2})\left[1-{2t}/({3t_0
 \beta})\right]$$
valid for $t\ll \left(\frac{R_0}{w}\right)^{1/2}$ and $t\ll t_{\rm
init}=\frac{2(\tilde{d}\eta_{\rm r}
 +\zeta_{\rm r})\beta}{\lambda v^2 m n_{cr}} \propto
 \frac{\sigma_0^2}{(\tilde{d}\eta_{\rm r}
 +\zeta_{\rm r})|\delta{\cal{T}}|}$.  Initial stage of the process
proceeds with  acceleration
$$ w=(d_{\rm sol}-1)\lambda
 v^2 \left(R_0 -R_{cr}\right)/(R_0 R_{cr}),$$
which changes sign at the initial size $R_0 =R_{cr}$ where
$$R_{cr}={(d_{\rm sol}-1)v^2\sqrt{2c\lambda}}/({3|\epsilon
|})\propto {1}/{|\delta{\cal{T}}|}$$ is the critical size. Germs
with $R_0 <R_{cr}$  shrink, while  germs with $R_0 >R_{cr}$ grow.
For germs with $|R_0 -R_{cr}|\ll R_{cr}$ the size changes very
slowly ($w\propto |\delta{\cal{T}}|(R_0-R_{cr})/R_{cr}^{2}$). For
undercritical germs of a small size, $w\propto
-|\delta{\cal{T}}|/R_0$. Slabs
of  stable phase, being placed in a metastable
medium, grow   independently of what was  their initial size. Note
that the same value $R_{cr}$ follows from minimization of the free
energy (\ref{Fsol}).

2) {\em{Long time evolution of a large germ.}} For $t \gg
t_{init}$, we may drop the term $ {\partial^2 \xi_0 }/{\partial
{\tau}^2}$ in the l.h.s of Eq. (\ref{ks}). For $R(t)\gg R_{cr}$,
surface effects become unimportant and we arrive at the solution
$$R(t) \simeq R_0 +u_{\rm asymp}t,\,\,u_{\rm asymp}=
{3|\epsilon|}\sqrt{{\beta}}/\sqrt{2\lambda {v^4}}.$$
Germs  grow with
 constant velocity.
The time scale for the growth of the germ with size $R\gg R_{cr}$
is
$t_{\rho}={R}/{u_{\rm asymp}}=
 ({m}/{T_{cr}})^{1/2}{R}/({(18\beta)^{1/2}|\delta{\cal{T}}|}).$
Asymptotic regime is reached  at very large values of  time,
provided the system is  near the critical point.

3) {\em{Long time evolution of a small germ.}} Describing germs of
a small size ($l\ll R \ll R_{cr}$, $d_{\rm sol}\neq 1$)   for
$t\gg t_{\rm init}$, we can drop the term $\propto
\widetilde{\epsilon}$ in (\ref{ks}). Then solution acquires the
form
$$ R(t)\simeq \sqrt{R_0^2  - 2(d_{\rm sol}-1){tl^2}/{t_0}}.$$
The time scale at which the  initial germ of a small size
dissolves is,
$t_{\rm dis}
=\frac{16n_{cr}T_{cr}(\tilde{d}\eta_{\rm r} +\zeta_{\rm
r})R_0^2}{(d_{\rm sol}-1)\sigma_0^2},$
and is  $\propto R^2_0$. Thus, fluctuations of sufficiently small
sizes are easily produced and  dissolve rapidly.

4) {\em{ Fluctuations in spinodal region.}} Let  the system be
driven to a spinodal region
where fluctuations of even   infinitesimally small amplitudes and
sizes may grow into a new phase. To demonstrate this we take the
free energy $\delta F$ to be close to its maximum ($\delta F
\simeq 0$). Then we  linearize Eq. (\ref{dimens}) dropping
$\psi^3$ term. Setting
$\psi=-\frac{\widetilde{\epsilon}}{2}+\mbox{Re}\{\psi_0
e^{\gamma_{\psi} \tau +i\vec{k}\vec{\xi}}\},$
  $\psi_0$ is an arbitrary but small real constant, we find
two solutions,
 \be\label{geng}
  \gamma_{\psi}(k)=(-k^2
  \pm \sqrt{{k^4}
+8{\beta}{k^2} -4{\beta}{k^4}}\,)/({2\beta}).
 \ee
 Growing modes correspond to the choice of "$+$"-sign  and $k^2 <2$.
  The time scale at which an
aerosol of germs develops is
$ t_{\rm aer}= t_0/\gamma_{\psi}(k_m),$
$k_{m}$ corresponds to $\mbox{max}\{\gamma_{\psi}(k)\}$. For an
effectively large viscosity ($\beta \ll 1$)  there are two
solutions: the  damped one, and  the growing one for $k<\sqrt{2}$.
The most rapidly growing mode is $\gamma_{\psi}(k_m )\simeq 2$,
$k_m =2\beta^{1/4}\ll 1$. The time scale characterizing
 growth of this mode is
$ t_{\rm aer}^{\eta}\sim \frac{1}{2}t_0
=\frac{4(\tilde{d}\eta_0 +\zeta_0)}{9n_{cr}T_{cr}|\delta
{\cal{T}}|}.$
The typical size   of  germs,
$ R_{\rm aer}^{\eta}\simeq l/(2\beta^{1/4}),$
 increases with an increase  of the viscosity.
For $k^2
>2$ both modes are damped.
In the case of an effectively small viscosity ($\beta \gg 1$) we
get
$\gamma_{\psi}(k)\simeq \pm k\sqrt{{2}/{\beta}} \sqrt{1-
{k^2}/{2}},$
 and
$\gamma_{\psi}^{max}(k_m =1)= \beta^{-1/2}.$ The time scale
characterizing  growing modes,
$  t_{\rm aer}^{\rm id}\sim t_0 /\gamma_{\psi}=
{2c^{1/2}}/{(\lambda v^2)}\propto {\delta{\cal{T}}}^{-1},$
  does not depend
on the viscosity in this limit.  The size scale of  germs is
$R_{\rm aer}^{\rm id}\simeq l.$
Modes with $k^2
>2$  oscillate and do not grow into
a stable phase.

 For the description of the hadron--QGP first-order phase
transition we take values $T_{cr} \simeq 162$~ MeV,
$n_{cr}/n_{sat}\simeq 1.3$, as they follow from lattice
calculations, see \cite{Aoki:2005vt}. Parameters of the EoS are
then as follows: $a\simeq 8.76\cdot 10^{2}(\mbox{MeV}\cdot
\mbox{fm}^{3})$, $b\simeq 1.60$~fm$^{-3}$, $\lambda \simeq
7.80\cdot 10^{-5}q^{-3} ({\mbox{fm}^{6}}/{\mbox{MeV}^{2}})$, $v^2
\simeq 1.56\cdot 10^{4}q^{2}|\delta {\cal{T}}|
({\mbox{MeV}^{2}}/{\mbox{fm}^{6}})$, $\epsilon \simeq 2.02\cdot
10^{2} (\delta {\cal{T}})^2 ({\mbox{MeV}}/{\mbox{fm}^{3}})$, where
$m$ is the effective quark mass, $q=({m}/{300\mbox{MeV}})$.
Further we obtain $l(T=0)\simeq 0.2$~fm (radius of confinement)
for $\sigma_0 \simeq 40$~MeV$/\mbox{fm}^2$. If one used $\sigma_0
\simeq 100$~MeV$/\mbox{fm}^2$, one would estimate  $l(T=0)\simeq
0.5$~fm.

 Next we use $s\simeq 7T^3
(T/T_{cr})$ at $T$ near $T_{cr}$,  $c_V \simeq 28 T^3 (T/T_{cr})$,
as it follows from the lattice data \cite{Aoki:2005vt}.  Assuming
the minimal value of the viscosity $\eta_{\rm min} =s/{4\pi}\simeq
60 $MeV/fm$^{2}$, $\zeta_{\rm min}=0$ we evaluate $\beta_{\rm
QGP}^{\rm max} \simeq 0.015q$ for $\sigma_0 \simeq
40$~MeV$/\mbox{fm}^2$, that corresponds to the limit of
{\em{effectively very large viscosity.}} Even for $\sigma_0 \simeq
100$~MeV$/\mbox{fm}^2$, $m =600$~MeV we would get $\beta_{\rm
QGP}^{\rm max} \simeq 0.2 \ll 1$. Note that following
\cite{Kharzeev} the bulk viscosity diverges in the critical point.
If were so ($\beta \rightarrow 0$), the quark-hadron system would
behave as absolutely viscous fluid, like glass, in near critical
region. Contrary, Ref. \cite{SR} argues for a smooth behavior of
the bulk viscosity.

With $\beta =0.015$
 we further estimate $t_0 \simeq 2 |\delta{\cal{T}}|^{-1}$~fm, $t_{\rho}\simeq 2.6R
q^{1/2}|\delta{\cal{T}}|^{-1},$ and $t_{\rm dis}\simeq 14 qR_0
\left({R_0}/{\mbox{fm}}\right)$.
 Typical time
 for the formation of the aerosol is
$t_{\rm aer}^{\eta}\simeq  |\delta{\cal{T}}|^{-1}
$~fm,  and typical size of  germs in aerosol is $R^{\eta}_{\rm
aer} \simeq 0.24|\delta{\cal{T}}|^{-1/2}
$~fm. Only $t_{\rm init}\simeq 0.03 q|\delta{\cal{T}}|^{-1}$~fm
 proves to be  small (excluding quite small $\delta{\cal{T}}$).
 Critical slowing down that limits growing of the
 $\sigma$ meson correlation length was discussed in
 \cite{Berdnikov}.

For the thermal conductivity we use an estimation $\kappa_{\rm
QGP} \simeq\alpha_0 \eta/m$ taking
 $\alpha_0 \simeq 3$ to recover the relation between values of $\kappa$ and
$\eta$ for nuclear gas-liquid phase transition at low energies
\cite{GIK}. The scale of the heat transport time is $t_T \simeq 26
q\left({R}/{\mbox{fm}}\right)^2$ fm.
 Using that $R_{cr}\simeq 0.1|\delta
{\cal{T}}|^{-1}$~fm, we obtain $R_{\rm fog} \simeq 0.1
q^{-1/2}|\delta{\cal{T}}|^{-1}~\mbox{fm} \lsim R_{cr}$.  The value
$R_{\rm fog}$ proved to be very small ($\lsim 0.1\div 1$~fm).
However,  time scale $t_T$ is rather long. Therefore, 
 the system most probably would have no time to fully  develop  a fog-like state in a
 hadron-quark phase transition in heavy ion collisions.

For the system in the vicinity of the critical point all estimated
time scales (except $t_{\rm init}$) are very large. If the system
trajectory paths   rather far from
 the critical point ($T_{cr}, \rho_{cr}$),
all time scales, except $t_T$, become less than the typical
life-time of the fireball
 ($\sim 10$~fm at RHIC conditions).
Reynolds numbers are $\lsim 1$, being  much smaller than the
critical value ($\gsim 1000$). Thereby, turbulence regime is not
reached.

We solved  numerically the general system of equations of nonideal
hydrodynamics for $d=2$. To illustrate the results we consider
dynamics of overcritical and undercritical germs (disks) in
infinite matter
 taking initial density profile as $\rho (x,y;t=0) = \rho_{out}
+ (\rho_{in}-\rho_{out}) \Theta({R_0}-r), \quad r=\sqrt{x^2+y^2}$,
$\rho_{in}$ and $\rho_{out}$ are  densities in  stable and
metastable homogeneous phases, respectively.
\begin{figure}
\centerline{%
{\includegraphics[height=3.5truecm] {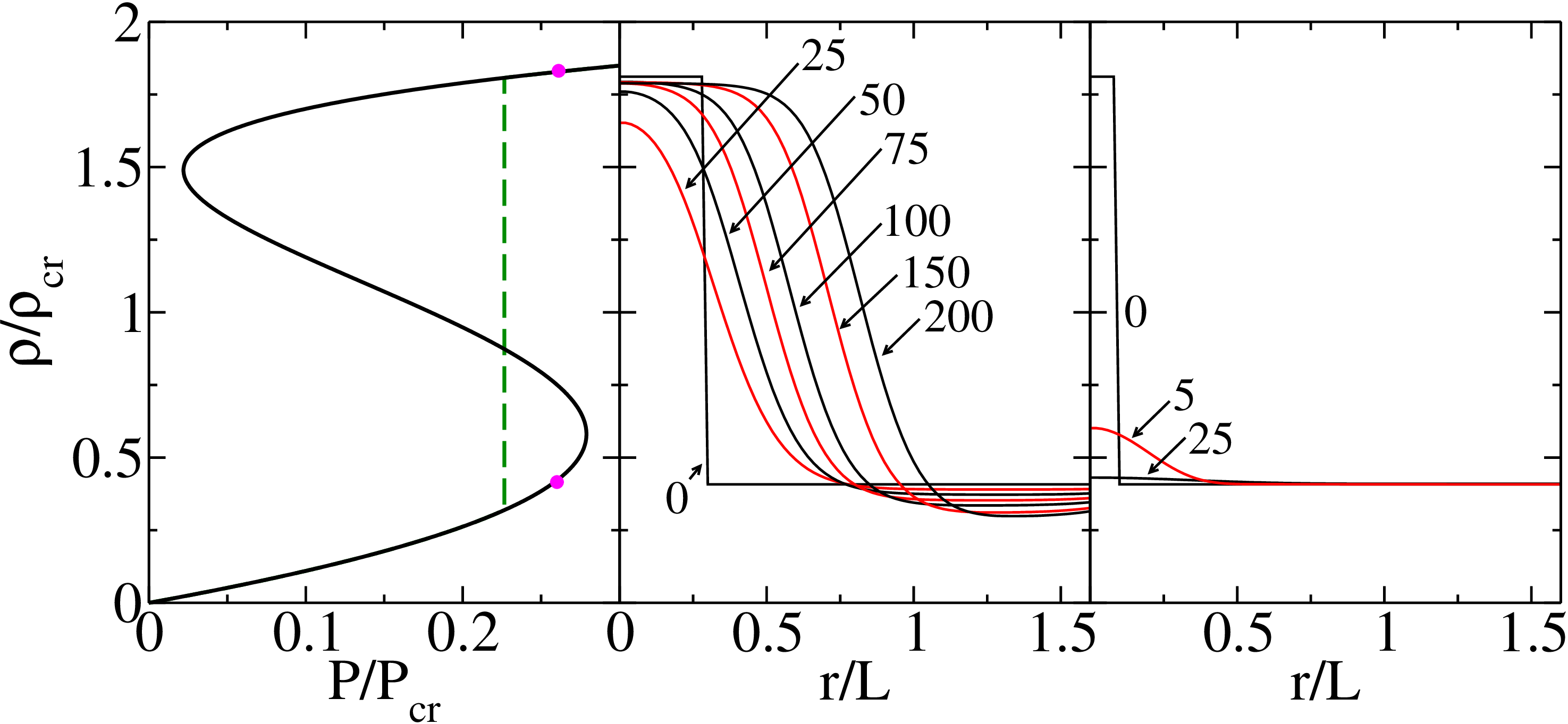}   } }
\vspace{5.0mm} \centerline{{\includegraphics[height=3.5truecm]
{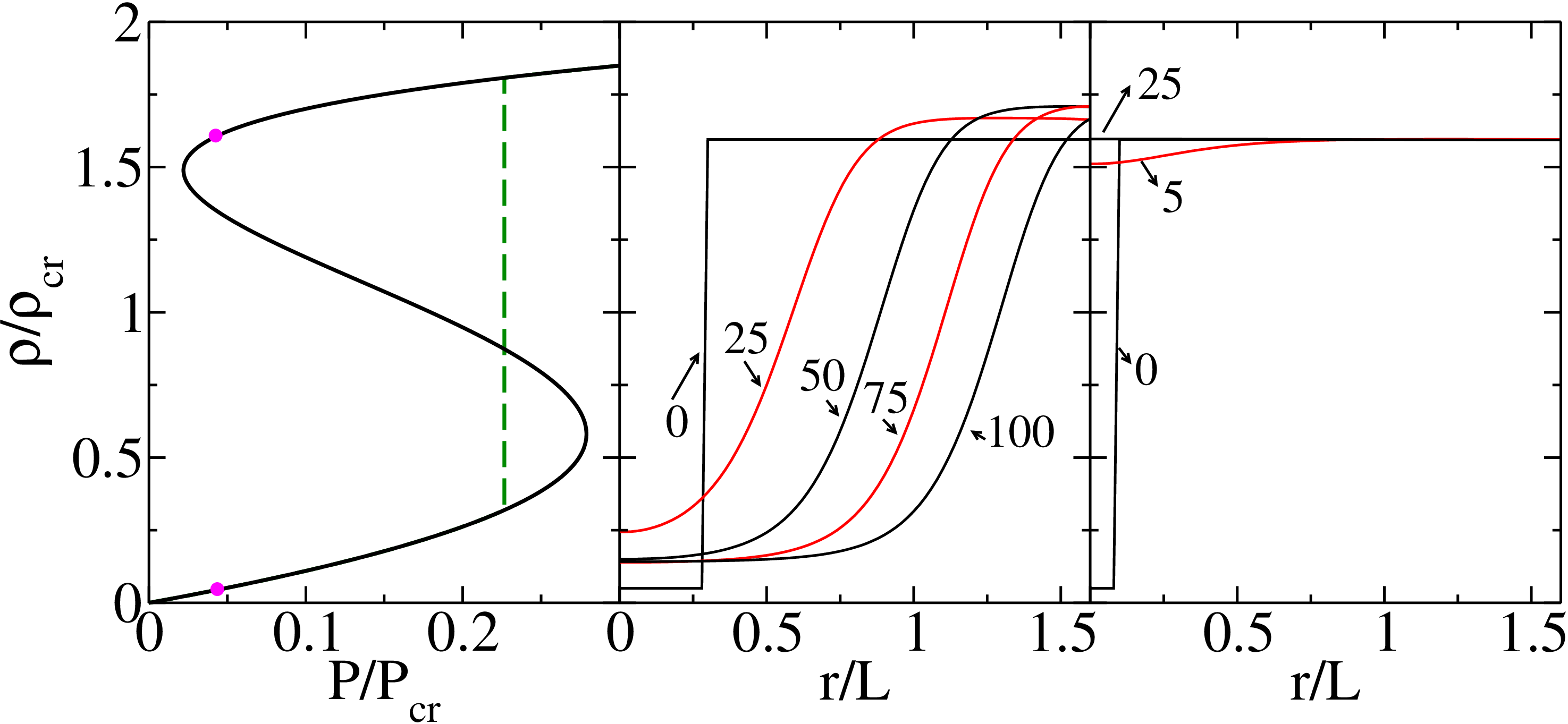} } } \caption{ Isotherm for the pressure as
function of the density with initial and final states shown by
dots  (left column).
 Dash vertical line shows the
Maxwell construction, MC.
In the upper panel the initial state corresponds to the stable
liquid phase disk in the metastable super-cooled gas, and in the
lower panel, to a stable gas phase disk in the metastable
super-heated liquid. Middle column shows time evolution of density
profiles for the
 overcritical liquid disk (upper panel) and gas disk (lower panel).
Numbers near curves (in $L$) are time snapshots.
Right column, the same for initially undercritical liquid or gas
disk.}\vspace{0.0mm} \label{EoS_dynamics}
\end{figure}

In Fig. \ref{EoS_dynamics} we  show the time evolution of a liquid
disk (upper panel) and a gas disk (lower panel) for $T/T_{cr}
=0.85$. In the middle column we show dynamics of an initially
overcritical germ with $R_0 =0.3L>R_{cr}\simeq 0.16L$ and in the
right column, of undercritical germ $R_0 =0.1L$, $L=5$~fm. The
 time
snapshots are shown in Figure in units L.
 The configuration is computed
for values of kinetic parameters  $\eta \simeq 45$~MeV$/$fm$^{2}$
and $\beta \simeq 0.2$.
 We see that in
case $R_0
> R_{cr}$ (middle column) disks slowly grow with the time passage.
For overcritical discs the initially selected distribution
acquires the $\mbox{tanh}$-like shape, see (\ref{sol}), only  for
$t\gsim (50\div 100)L$. Initial disks of a small size practically
disappear for $t\gsim 10L= 50$~fm.
 Due to  the matter
admission  to the disk surface and the shape reconstruction, the
density decreases in the liquid disk neighborhood  below  the
value of the density in the homogeneous metastable matter and it
increases in the gas disk surrounding above the value of the
density in the homogeneous metastable matter (see the middle
column).

\begin{figure}\vspace{6mm}
\centerline{%
\includegraphics[height=3.0truecm]{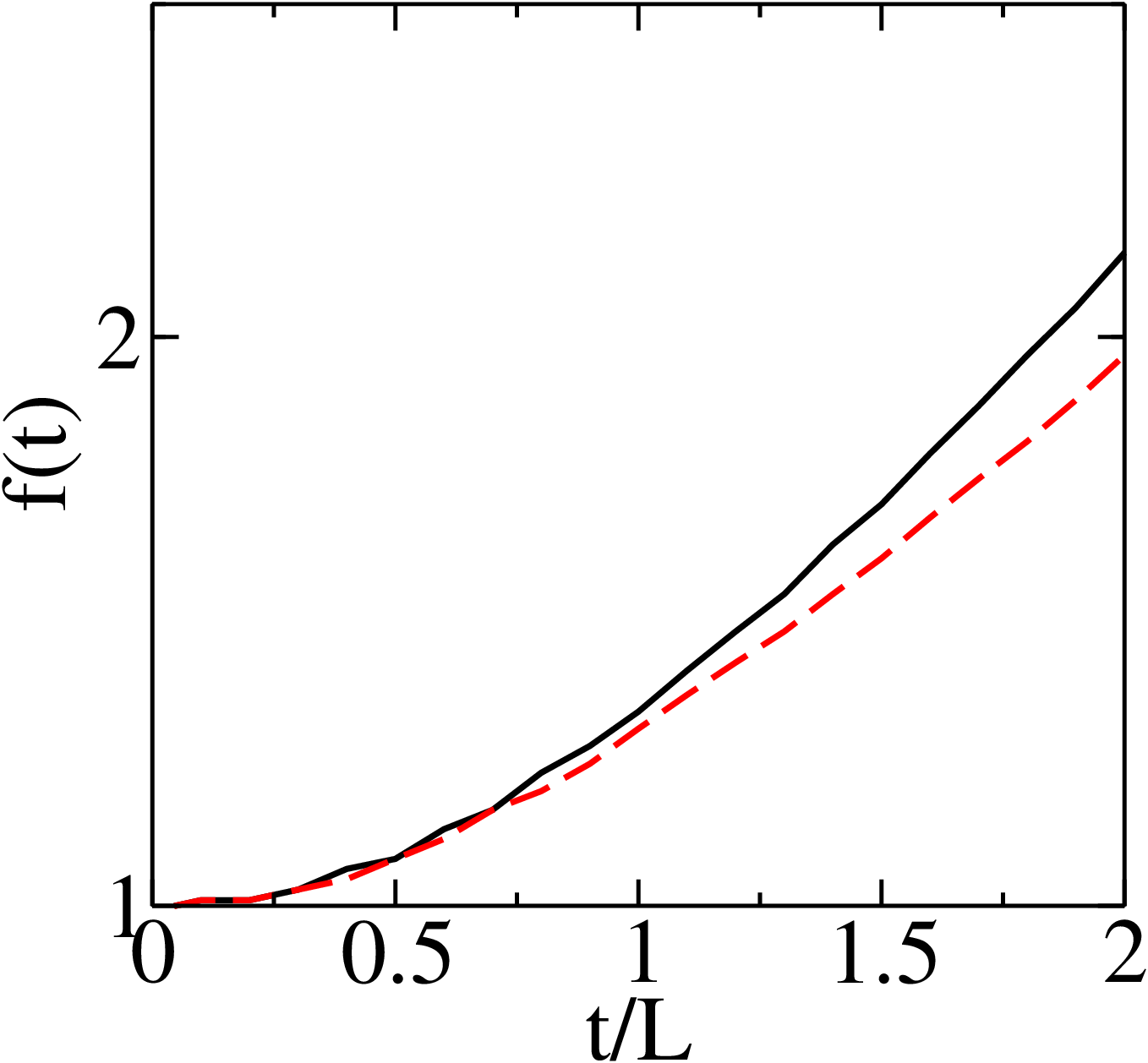}
\hspace{-1.0mm}
\includegraphics[height=3.0truecm] {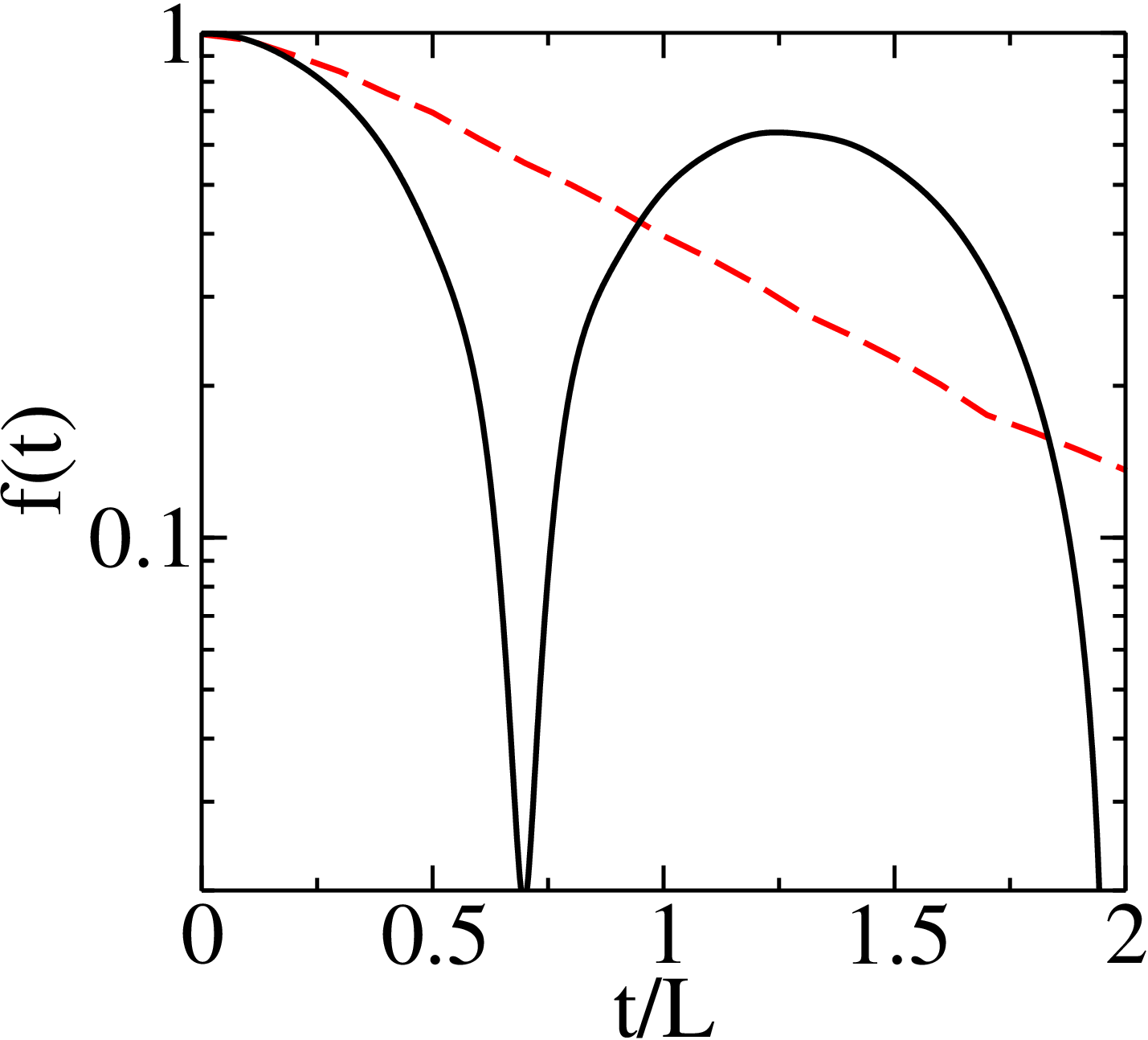} }
\caption{ Time evolution of  wave amplitudes $f(t)$ in aerosol for
effectively small ($\beta =20$, solid line) and  large  ($\beta
=0.2$, dash line) viscosity. Left panel:
$k=2l/L$ (growing modes).
Right:
$k=8l/L$ (oscillation modes for large $\beta$ and damping modes
for small $\beta$).  Other parameters are the same as in Fig
\ref{EoS_dynamics}.
}\label{waveampl}
\end{figure}

In Fig. \ref{waveampl}  we demonstrate time evolution of the wave
amplitudes, $\rho (t)=\bar{\rho}+A_0 f(t)\mbox{sin}
(\vec{k}\vec{r})$, for an undercritical value of the wave number
$k$ (left panel) and for an overcritical value (right panel).
 In case of the overcritical  $k$ and effectively small viscosity
($\beta =20$) we demonstrate change of the amplitude in the
$3/2$-periods of the oscillation. Such a behavior
 fully
agrees with that follows from our analytical treatment of the
problem.
 
 Concluding, even in the spinodal region germs are
growing  slowly, if the system is  somewhere in the vicinity of
the critical point. Thus in heavy ion collisions the expanding
fireball may linger in the QGP state, until $T(t)$ decreases below
the corresponding equilibrium value of the temperature of the
phase transition. There exists a belief that strongly coupled QGP
state, represents almost perfect fluid \cite{Romatschke:2007mq}.
We demonstrate {\em{the essential role of  viscosity and surface
tension  in dynamics of  first-order phase transitions, including
the hadron-QGP one.}} Fluctuations in QGP (at a finite baryon
density) grow and dissolve as if the fluid were very viscous.
Variation of parameters in broad limits does not change
conclusions.

%\vspace*{5mm} {\bf Acknowledgements} \vspace*{5mm}

We are grateful to  B. Friman, Y.B. Ivanov, E.E. Kolomeitsev, J.
Randrup, and V.D. Toneev  for numerous  discussions. This work was
supported in part by the
 DFG
project 436 RUS 113/558/0-3, and
RFBR grants 06-02-04001 and 08-02-01003-a.

\end{document}